\documentclass[draft]{agujournal2019}
\usepackage{url} 
\usepackage{lineno}
\usepackage[inline]{trackchanges} 
\usepackage{soul}
\usepackage{float}

\usepackage{array}

\draftfalse

\journalname{Journal of Advances in Modeling Earth Systems (JAMES)}

\begin{document}

%
%


\title{Towards a Unified Data-Driven Boundary Layer Momentum Flux Parameterization for Ocean and Atmosphere}

%
%




\authors{Renaud Falga\affil{1}, Sara Shamekh\affil{1}, Laure Zanna\affil{1}}


\affiliation{1}{Courant Institute of Mathematical Sciences, New York University, New York, NY, USA}




\correspondingauthor{Renaud Falga}{renaudfalga@gmail.com}



\begin{keypoints}
\item A data-driven parameterization for both atmospheric and oceanic boundary layer momentum flux is developed and implemented in SCAM.
\item The neural network is able to predict upgradient fluxes, highlighting the potential to outperform traditional eddy diffusivity schemes.
\item The ANN parameterization outperforms the CLUBB scheme in replicating the LES wind profiles, especially in convective regimes.
\end{keypoints}

%
%

%
%


\begin{abstract}
Boundary layer turbulence, particularly the vertical fluxes of momentum, shapes the evolution of winds and currents and plays a critical role in weather, climate, and biogeochemical processes. In this work, a unified, data-driven parameterization of turbulent momentum fluxes is introduced for both the oceanic and atmospheric convective boundary layers. An artificial neural network (ANN) is trained offline on coarse-grained large-eddy simulation (LES) data representing a wide range of turbulent regimes in both fluids. By normalizing momentum flux profiles with their surface values, we exploit a self-similar structure across regimes and fluids, enabling joint training. The ANN learns to predict vertical profiles of subgrid momentum fluxes from mean wind or current profiles, capturing key physical features such as upgradient fluxes that are inaccessible to traditional first-order closure schemes. When implemented online in the Single Column Atmospheric Model (SCAM), the ANN parameterization consistently outperforms the SCAM baseline parameterization in replicating the evolution of the boundary layer wind profiles from the LES, especially under convective conditions, with errors reduced by a factor of 2–3 across regimes. ANN performance remains robust even when the surface momentum flux is biased by ±30\%, and generalization is confirmed by testing on LES cases excluded from the training dataset. This work demonstrates the potential of machine learning to create unified and physically consistent parameterizations across boundary layer systems in climate models.
\end{abstract}

\section*{Plain Language Summary}
In both oceanic and atmospheric boundary layers, the transfer of momentum that is induced by turbulence has a significant impact on winds, currents, and many large-scale climatic phenomena. It is therefore essential for coarse-resolution models to accurately parameterize the subgrid effects of turbulence on circulation patterns, yet current methods still struggle to represent the fluxes of momentum. In this study, we introduce a novel machine learning approach that utilizes a neural network trained on high-resolution simulations to predict the subgrid fluxes of momentum in both the atmosphere and the ocean. By uncovering a common structure in turbulent behavior, the same model can be applied to both systems. When tested in a single-column climate model framework, the neural network outperforms traditional methods, especially during convective conditions. This unified approach offers a promising path to improve weather and climate predictions.

\section{Introduction}
The oceanic and atmospheric boundary layers exhibit important physical similarities. In both, the transport of heat, momentum, and tracers is dominated by turbulence, which can be generated either by shear or by surface radiative forcing that generates buoyancy. Boundary layers are typically classified as stable, neutral, or convective, depending on their stratification. Convective boundary layers (CBLs), specifically, share particular similarities across fluids and can form well-mixed layers with nearly uniform temperature, winds, or currents. Although the nature of the surface forcing differs — with convection driven by daytime heating in the atmosphere and by nighttime cooling or precipitation in the ocean — the resulting turbulent dynamics and vertical structure are comparable. In both fluids, vertical turbulent fluxes such as momentum and heat peak at the surface and decrease throughout the boundary layer (BL), reflecting analogous energy transfer processes.

        Because of poor representation of boundary layer processes, numerical weather forecasts and climate models struggle to accurately simulate near-surface atmospheric variables like temperature and wind \cite{edwardsAssessmentNumericalWeather2011,beljaarsStableBoundaryLayer2011}, or oceanic mixed layer depth and sea-surface temperature \cite{belcherGlobalPerspectiveLangmuir2012}. In the atmosphere, different biases in temperature and moisture can arise from the choice of BL parameterization, both in stable and convective conditions \cite{huEvaluationThreePlanetary2010}. The biases in wind, temperature and clouds also induce important biases in near-surface oceanic variables including sea surface temperature (SST) in coupled climate runs \cite{richterClimateModelBiases2015}. Specifically, these positive SST biases are partly caused by unrealistically low stratocumulus cloud cover \cite{huLowCloudErrors2008} and an underestimated near-surface wind speed that reduces upwelling \cite{vanniereSystematicApproachIdentify2014}, which highlights the complex interactions between the atmospheric and oceanic boundary layers. The biases mentioned above could be reduced by improving the representation of boundary layer turbulence in coarse resolution models.

            Among the turbulent fluxes, the fluxes of momentum play a central role in shaping the vertical structure of wind and current profiles. In global climate models (GCMs), BL turbulence needs to be parameterized, and accurately representing momentum fluxes is critical for reliable weather and climate projections. The misrepresentation of atmospheric momentum fluxes has been shown to lead to biases in surface wind patterns and resulting Hadley circulation in GCM simulations \cite{richterEffectsConvectiveMomentum2008}, which can in turn affect the cloud cover. Boundary layer clouds are commonly poorly simulated in GCMs \cite{teixeiraTropicalSubtropicalCloud2011}, and the impact of clouds on future climate is one of the largest sources of uncertainty within climate model projections \cite{caldwellQuantifyingSourcesIntermodel2016}. Accurate parameterization of atmospheric boundary layer (ABL) momentum flux has been shown to improve the representation of clouds in GCMs \cite{suseljImprovingRepresentationSubtropical2021}. Momentum fluxes are key critical processes in the oceanic boundary layer (OBL) as well. It is estimated that 90\% of the excess heat induced by anthropogenic greenhouse gases is absorbed by the ocean \cite{resplandyQuantificationOceanHeat2019}, and the strength of convectively-driven mixing partly determines the amount of heat and carbon uptake. Improving representation of oceanic boundary layer momentum fluxes actually leads to an improvement in the simulation of global ocean heat uptake \cite{shimuraImpactsOceanWaveDependent2020}. Given their impact on surface exchanges as well as large-scale atmospheric and oceanic processes, it is essential to develop accurate parameterizations of momentum fluxes for GCMs.

    Among the most common schemes that have been adopted in ABL and OBL parameterizations are the first-order eddy-diffusivity schemes. In such models, the momentum flux is expressed as the opposite of the vertical gradient of mean wind (or current) multiplied by an eddy-diffusivity coefficient; the flux is thus said to be downgradient. However, these schemes usually fail to represent convective boundary layers, because of the well-mixed structure of the wind and current profiles.  \cite{brownErrorsParametrizationsConvective2006}. Additionally, in reality, upgradient fluxes can occur because of non-local transport or mesoscale dynamics, and therefore cannot be represented by these approaches \cite{dixitCounterGradientMomentumTransport2021,helferRoleShallowConvection2021}. Some schemes added a mass-flux term to represent the effect of non-local mixing on turbulent transport. However, it has been shown that these so-called mass-flux schemes fail to represent the vertical momentum transport for both shallow and deep convection \cite{zhuMassFluxRepresentationVertical2015}. The representation of wind and boundary layer clouds can be improved by higher order schemes that add prognostic equations for the momentum fluxes \cite{larsonMomentumTransportShallow2019,nardiAssessingSensitivityTropical2022,graapUsingEUREC4ATOMIC2024}. However, these approaches remain computationally demanding and add complexity to the models by introducing additional prognostic equations, tunable parameters, and higher order terms. Climate models and operational forecasts can be tuned in order to match observations or high resolution simulations. For instance, artificially reducing the mixing in the stable BL has been shown to improve the representation of surface wind, however, it can also deteriorate large-scale flow and near-surface temperature \cite{Sandu2013}. \citeA{cuxartSingleColumnModelIntercomparison2006} also highlighted the sensitivity of the models to the different closure parameters, and that the simulation accuracy was influenced more by parameter tuning than by the choice of closure itself. Therefore, increasing the number of tunable parameters might increase the complexity of the models, while reducing interpretability.

   In recent years, machine learning–based parameterizations have emerged as a rapidly developing alternative to traditional approaches. Neural networks have been used to predict the ABL subgrid turbulent fluxes \cite{shamekhLearningAtmosphericBoundary2023}, subgrid-scale stresses \cite{connollyDeepLearningTurbulence2025} and deep convective momentum fluxes \cite{yuvalNeuralNetworkParameterizationSubgrid2023}. Machine learning methods have also been used to improve existing eddy-diffusivity mass-flux (EDMF) schemes by tuning the remaining unknown parameters of the scheme such as entrainment rate, in an online setup \cite{christopoulosOnlineLearningEntrainment2024}. In the ocean, data-driven approaches have been developed to improve existing OBL schemes \cite{ramadhanCapturingMissingPhysics2023,saneParameterizingVerticalMixing2023}, and parameterize the effects of mesoscale eddies \cite{zannaDataDrivenEquationDiscovery2020,guillauminStochasticDeepLearningParameterization2021,zhangImplementationEvaluationMachine2023,perezhoginGeneralizableNeuralnetworkParameterization2025}. Yet, no efforts have been made to replace existing boundary layer momentum flux parameterization by a machine learning-predicted flux. Furthermore, despite the numerous similarities between the ABL and OBL dynamics — and their common numerical representations within GCMs — a unified parameterization applicable to both BLs has yet to be developed. Such an approach could provide a novel, physically consistent framework for both BLs, while also bridging a gap between atmospheric and oceanic research communities. Machine learning methods can also benefit from the multi-fluid framework by increasing the amount of training data, provided that a proper normalization procedure is applied to make atmospheric and oceanic features comparable. Here for the first time, we develop a unified data-driven parameterization that predicts vertical profiles of momentum fluxes in both atmospheric and oceanic CBLs. The parameterization is implemented in the single-column atmospheric model or SCAM \cite{gettelmanSingleColumnAtmosphere2019} and tested in idealized configurations. In section 2, the problem is formulated and the methods used to train the neural network and test it online are presented. The offline prediction skills of the artificial neural network (ANN) are then presented in section 3. In section 4, we demonstrate that the ANN parameterization implemented in SCAM consistently outperforms the existing Cloud Layers Unified by Binormals (CLUBB), which is a first order scheme for momentum fluxes, in idealized configurations. Sections 5 presents a conclusion and discussion of the results, with potential improvements and next steps to make this new data-driven approach operational in more realistic situations.

\section{Methods}
\subsection{Problem formulation}
In the boundary layer, the Reynolds-averaged Navier-Stokes horizontal momentum equations can be expressed as followed \cite{Stull1988}:

\(
\frac{\partial \overline{u}}{\partial t} = -\mathbf{\overline{u}} \cdot \nabla \overline{u} -f\overline{v} - \frac{1}{\rho}\frac{\partial p}{\partial x} - \frac{1}{\rho} \frac{\partial \rho \overline{u'w'}}{\partial z} + \nu \frac{\partial^2\overline{u}}{\partial x^2}
\)  (1),

\(
\frac{\partial \overline{v}}{\partial t} = -\mathbf{\overline{v}} \cdot \nabla \overline{v}  +f\overline{u}- \frac{1}{\rho}\frac{\partial p}{\partial y} - \frac{1}{\rho} \frac{\partial \rho\overline{v'w'}}{ \partial z} + \nu \frac{\partial^2\overline{v}}{\partial y^2}
\)  (2),

where $u$, $v$ and $w$ are respectively the zonal, meridional and vertical wind velocities, p is the pressure, f the Coriolis parameter, $\rho$ the air density, $\overline{u'w'}$ and $\overline{v'w'}$ the turbulent fluxes of horizontal momentum and $\nu$ the kinematic viscosity. The common Reynolds notations are used here, where $\overline{x}$ represent the spatial mean of variable $x$, and $x'$ represents the deviation to the mean, or perturbation. The five terms on the right hand side of both equations represent respectively the advection of mean wind, the effect of the Coriolis force, the effect of horizontal pressure gradients, the vertical turbulence terms and the influence of horizontal viscous stress. In GCMs, the effects of fluid-specific physical processes (e.g., wave breaking or Langmuir circulation in the ocean, condensation in the atmosphere) are included in the subgrid turbulent term. Therefore, the momentum equations are identical in both atmospheric and oceanic boundary layers. In the atmosphere, molecular diffusion is several orders of magnitude weaker than turbulent mixing, and the viscous terms are therefore usually neglected. 

    In GCMs, the components of the momentum flux $\overline{u'w'}$ and $\overline{v'w'}$ have to be parameterized in both boundary layers. In the first-order eddy-diffusivity approach, the terms are parameterized following:

\( \overline{u'w'} = -K \frac{\partial \overline{u}}{\partial z} \)    (3),

\( \overline{v'w'} = -K \frac{\partial \overline{v}}{\partial z} \)    (4),

where $K$ is an eddy-diffusivity coefficient which varies with height and can depend on different physical factors such as the turbulent kinetic energy (TKE), or a mixing length scale \cite{hotslag1993Local,Mellor1974,Bougeault1986,Golaz2002,troen1986simple,priestley1947vertical,Han2019,lopez2020generalized}. When a prognostic equation is used for the TKE, the scheme is then said to be 1.5 order. Differing from the traditional methods, a data-driven approach is adopted in this study, in which an ANN is used to predict the vertical profiles of atmospheric (oceanic) momentum flux, given the vertical profiles of mean wind (current) components. The ANN is trained offline on coarse-grained high resolution idealized LES representing various turbulent convective regimes. Two different LES models were used to generate samples from both ocean and atmosphere. The trained ANN is then tested online after implementation in the Single Column Atmospheric Model or SCAM, and the ANN skill is compared to the existing eddy diffusivity scheme CLUBB. Note that CLUBB is a higher order scheme for most BL turbulent fluxes, but still uses a 1.5 order approach to parameterize the momentum fluxes, in its current formulation. It's also worth noting that the CLUBB momentum flux is planned to be upgraded to prognostic formulations in the next version of CAM. Subsection 2.2 describes the methodology used to train the ANN, which includes the descriptions of the LES setup as well as the ANN parameters. The single column model framework and the baseline momentum flux parameterization are then described in subsection 2.3.

\subsection{ANN training from coarse-grained high-resolution simulations}
\subsubsection{Large-eddy simulations}
High-resolution simulations of dry atmospheric BL ranging from neutral to highly convective conditions were performed using the LES code MicroHH \cite{vanheerwaardenMicroHH10Computational2017}. Simulations were ran under idealized conditions, with constant surface heating and geostrophic wind forcing. The model domain has 256 grid cells in both horizontal direction with a resolution of twenty-four meters, covering a distance of 6,120 meters. There are 384 vertical levels with a resolution of six meters, the top of the domain reaching 2,301 meters of altitude. A sponge layer is included at the top in order to prevent the reflection of upward-propagating waves. The domain is doubly periodic in the horizontal direction. The initial potential temperature is set to be constant at 300 K below the boundary layer top (set at an initial 800 m), and increases with a rate of 5 K/km above. We run fifteen simulations with different combinations of geostrophic wind $U_g$ and surface heat flux forcing Q, spanning a wide range of turbulent regimes: from quasi-neutral to highly convective. The turbulent regimes are quantified using the stability parameter $\zeta = -z_i/L$, where $z_i$ is the PBL height and L is the Monin-Obukhov length, which is defined as L = $u_{*}^3 .\overline{\theta} / (kg\overline{w'\theta '})$, with $u_*$ being the friction velocity, $\theta$ the potential temperature, $k$ the Von Karman constant, and w the vertical velocity. Low values of $\zeta$ represent quasi-neutral regimes where the turbulence is mainly shear-generated, while high values of  $\zeta$ correspond convectively-driven turbulent regimes. 

    Oceanic high-resolution simulations were conducted using the ocean modeling software Oceananigans \cite{wagnerHighlevelHighresolutionOcean2025}. OBL convection simulations were obtained by applying a constant surface cooling and wind forcing. The model domain has 256 grid cells with a 2 meter grid spacing in both horizontal directions. The vertical axis has 128 grid cells with a 2 meter grid spacing, starting at one meter below the surface and reaching a depth of 255 meters. Similarly as in  \citeA{ramadhanCapturingMissingPhysics2023}, the domain is initialized with a stratified fluid at rest with weak stratification in the surface layer (initially 50 meters deep) and the deep ocean, separated by a strongly stratified thermocline with a width of 25 meters. At the beginning of each simulation, surface cooling is applied via a negative surface buoyancy flux, and wind forcing is applied to generate mixing and deepening of the OBL. Both forcing are applied and remain constant throughout the whole simulation length. Twelve oceanic LES simulations are ran with different combination of surface wind stress and cooling. Table 1 and 2 in the Appendix summarize the atmospheric and oceanic LES simulations settings and their corresponding stability parameter.

\subsubsection{Neural network training}
    In order to train the neural network to predict subgrid momentum fluxes, a coarse-graining procedure is applied to the high-resolution data. This process involves averaging the data horizontally, effectively reducing the resolution from a fine 256x256 grid to a coarser 2x2 grid. Four coarse-grained samples (i.e., vertical profiles) of each variable are thus obtained after coarse-graining, at each time step. The subgrid momentum fluxes are then computed following:
    $\overline{x'w'} = \overline{xw} - \overline{x}$ $\overline{w}$, where $x$ can be either one of the horizontal wind components. For atmospheric LES runs, the simulations are ran for two hours and an average of 395 samples per simulation are kept for training, resulting in 5,920 samples. The ocean LES simulations are ran for four days and 334 samples per simulation are kept on average, resulting in 4,004 samples. Considering both fluids, this results in a total of 9,924 samples.

    An ANN with two hidden layers, each containing 128 neurons, is trained to predict both vertical profiles of the turbulent momentum fluxes $\overline{u'w'}$ and $\overline{v'w'}$. In order to prevent overfitting, a 0.2 dropout is applied at each layer, as well as a $10^{-3}$ weight decay. Two different training scenarios are used for offline and online. For offline predictions, vertical profiles of  $\overline{u}$, $\overline{v}$ and TKE are used as inputs. For both fluids, the vertical profiles of inputs and outputs are interpolated on the normalized $z$ axis $z/z_i$, ranging from $z/z_i = 0.05$ to $z/z_i=1.20$, with a vertical spacing of $\Delta z=0.02$ (2\% of $z_i$), resulting in 58 vertical levels. For the ANN implemented online, a simplified scenario is used. Only the vertical profiles of $\overline{u}$ and $\overline{v}$ are used as inputs, and the ANN is trained on the resolution of the host model SCAM (additional details are provided in the next subsection). This simplified setup prevents any bias arising from the prediction of TKE and boundary layer height by the host model.

\begin{figure}[ht]
    \centering
    \includegraphics[width=1\linewidth]{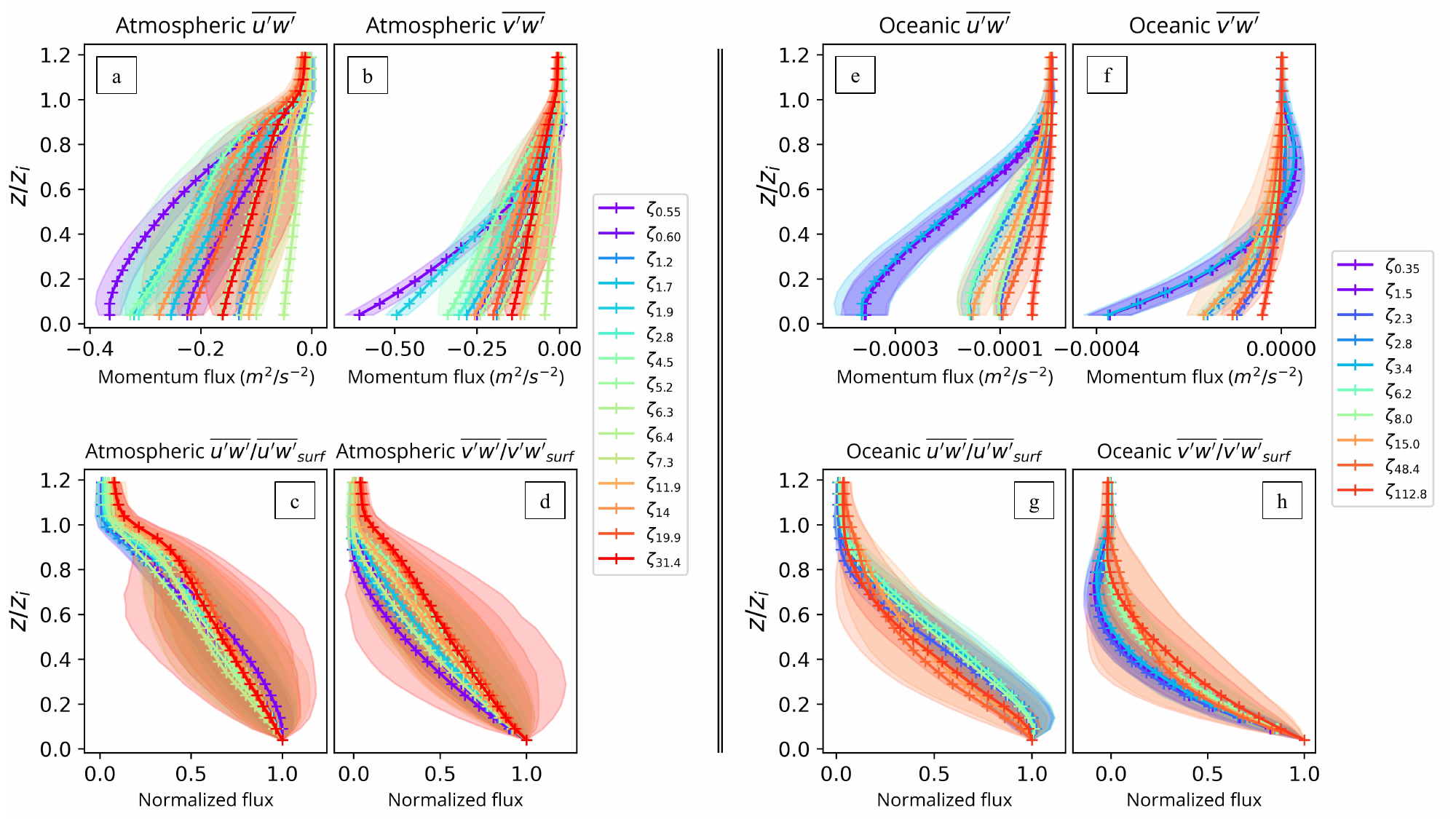}
    \caption{Vertical profiles of turbulent momentum fluxes computed from the atmospheric LES (a and b) and oceanic LES (e and f). The fluxes normalized by their surface value are shown in panels c, d, g and h. The common vertical levels used for joint training are labeled as vertical level indices. Each thick line represents the average flux from a single LES simulation, noted as $\zeta_x$, with $x$ being the stability parameter of the corresponding run. The shaded area corresponds to the flux standard deviation of the simulation. The atmospheric vertical profiles have been interpolated on the SCAM momentum levels. The oceanic vertical profiles have been interpolated on the same number of vertical levels, evenly spaced. The ocean fluxes have also been flipped in the normalization process in order to match the shape of the atmospheric fluxes.}
    \label{fig:1}
\end{figure}

    Neural networks typically perform poorly when predicting data outside the training range. To mitigate the risk of out-of-distribution predictions, the normalization procedures were carefully designed. Aligning distributions through statistical or physical transformations has been shown to enhance the generalizability of neural networks \cite{beuclerClimateinvariantMachineLearning2024,perezhoginGeneralizableNeuralnetworkParameterization2025,connollyDeepLearningTurbulence2025}. Therefore, normalization was implemented to ensure that the input and output distributions remained consistent across all of the different turbulent regimes. Each vertical profile of $\overline{u'w'}$ and $\overline{v'w'}$ is normalized by its surface values, and the resulting dimensionless fluxes $\overline{u'w'}/{\overline{u'w'}_{surf}}$ and $\overline{v'w'}/{\overline{v'w'}_{surf}}$ exhibit self-similarity across turbulent regimes, appearing nearly identical regardless of the conditions (Fig. \ref{fig:1}). The momentum fluxes show close resemblance in atmospheric and oceanic BLs. In the atmosphere, horizontal momentum is on average transferred downward, from the high quasi-geostrophic wind speeds of the free troposphere to the surface, where the wind is weaker because of surface friction. The momentum from the wind stress at the ocean surface is also transferred downward, from the surface to the deep ocean. For both fluids, this results in a vertical profile of momentum flux that is maximum in magnitude at the surface, and progressively decreasing quasi-linearly until the limit of the BL. It's important to note that in GCMs, the surface fluxes are usually computed outside of the BL parameterization by the surface model following Monin-Obukhov theory, and given as inputs to the atmospheric and oceanic boundary layer schemes \cite{moninBasicLawsTurbulent}, which makes the reverse normalization procedure achievable in practice within a GCM. The same type of normalization is used for the TKE offline, and each vertical profile of TKE is divided by its surface value. The vertical profiles of the inputs $\overline{u}$ and $\overline{v}$, on the other hand, are normalized using a statistical approach, by subtracting the mean and dividing by the standard deviation of each vertical profile.

\subsection{Online implementation in SCAM}
    SCAM \cite{gettelmanSingleColumnAtmosphere2019} is a single column model version of the Community Atmosphere Model (CAM), the atmospheric component of the coupled Community Earth System Model (CESM). It predicts temperature, humidity and momentum within a column using large-scale (forcing terms) and local tendencies (parameterized physics), using the same physics and same parameterizations as CAM. In this study, SCAM is used to replicate the spatially averaged BL wind profiles obtained from the fifteen atmospheric LES.

    To train the implemented ANN, both atmospheric inputs and outputs are interpolated on the SCAM vertical resolution. In its current implementation, SCAM only allows for a single vertical resolution. The wind components are computed on model levels, while momentum fluxes are computed on interface levels. Therefore, before training, the $\overline{u}$ and $\overline{v}$ LES vertical profiles are interpolated on the SCAM model levels falling within the vertical extent of the LES domain (eight levels) and the vertical profiles of $\overline{u'w'}$ and $\overline{v'w'}$ on the SCAM interface levels (nine levels, see Fig. \ref{fig:1}). In this study, the neural network is trained on the same resolution as the host model, and the sensitivity of the ANN to different host model resolution is not treated. Note that the ANN trained for implementation in SCAM still includes the oceanic features. For the vertical resolution of those features, we also interpolate  $\overline{u}$ and $\overline{v}$ ($\overline{u'w'}$ and $\overline{v'w'}$) on eight (nine) vertical levels evenly spaced located between -3 and -155 meters depth.

    The trained ANNs are imported from PyTorch (Paszke et al, 2019) to the SCAM Fortran code using the FTorch library \cite{atkinsonFTorchLibraryCoupling2025}. To replicate the LES runs using SCAM, the model is initialized with the initial LES vertical profiles of $u$, $v$, and potential temperature $\theta$ (averaged horizontally across the LES domain). The same forcing conditions (geostrophic wind and surface heating) are also used. The large-scale advection of mean wind (equation (1) and (2), first term) is neglected here, consistent with the doubly periodic LES domain. By replacing the horizontal pressure gradients term with the geostrophic wind ($u_g, v_g$)  and neglecting viscosity, equations (1) and (2) thus simplify to:

\(
\frac{\partial \overline{u}}{\partial t} = -f(v_g-\overline{v})  - \frac{1}{\rho} \frac{\partial \rho \overline{u'w'}}{\partial z} 
\)  (1b), 

\(
\frac{\partial \overline{v}}{\partial t} = +f(u_g-\overline{u})  - \frac{1}{\rho} \frac{\partial \rho \overline{u'w'}}{\partial z} 
\)  (2b).

    In such setup, the temporal evolution of the wind profiles is thus only affected by the Coriolis force and turbulence. Therefore, by comparing the wind profile temporal evolution predicted by SCAM with the LES wind profiles, the turbulence parameterization accuracy is assessed. Furthermore, to test whether the ANN parameterization can perform well on unseen turbulent regimes, we train a new ANN excluding one LES simulation, and use SCAM to reproduce the vertical profile evolution of that excluded case. Finally, the performance of the ANN parameterization is compared to the CLUBB scheme. In the CLUBB runs, we also initialize the TKE using the profiles obtained after two hours of spin-up. Therefore, for each one of the fifteen LES runs, three SCAM simulations are conducted:

     \begin{itemize}
         \item One with the ANN trained on all of the LES runs.
        \item One with an ANN trained after excluding the LES run from the training process.
        \item One with CLUBB (baseline).
    \end{itemize}

This yields a total of 45 SCAM simulations. In coupled GCMs, the surface values are inputs to the BL schemes, therefore, the LES surface values are prescribed in all of the SCAM runs, at each time step. Note that the surface values are prescribed at each of the CLUBB substep instead of the SCAM time step, which has been shown to improve representation of surface winds \cite{gentileEffectCouplingCLUBB2024}. Although in this case, values are prescribed and not computed using bulk formulas. In both cases however, the BL scheme prediction partly depends on the prediction of the surface fluxes by other components of the model. In the case of CAM, it has been shown that the simulated surface momentum fluxes matched observations during daytime convective BL quite well, but could slightly underestimate them during nighttime \cite{burnsComparisonDielCycle2018}. Friction velocities can also be overestimated when an additional turbulent mountain stress is included \cite{lindvallEvaluationNearSurfaceParameters2013}. Therefore, in order to assess the different parameterizations sensitivity to the predicted surface values, additional sensitivity analyses were performed. For each one of the above-mentioned runs, six additional SCAM simulations were conducted after applying biases to the prescribed surface momentum flux values (-30\%, -20\%, -10\%, +10\%, +20\% and +30\%).

\section{Offline prediction}
\subsection{Prediction of a random test set}
\begin{figure}[h]
    \centering
    \includegraphics[width=1\linewidth]{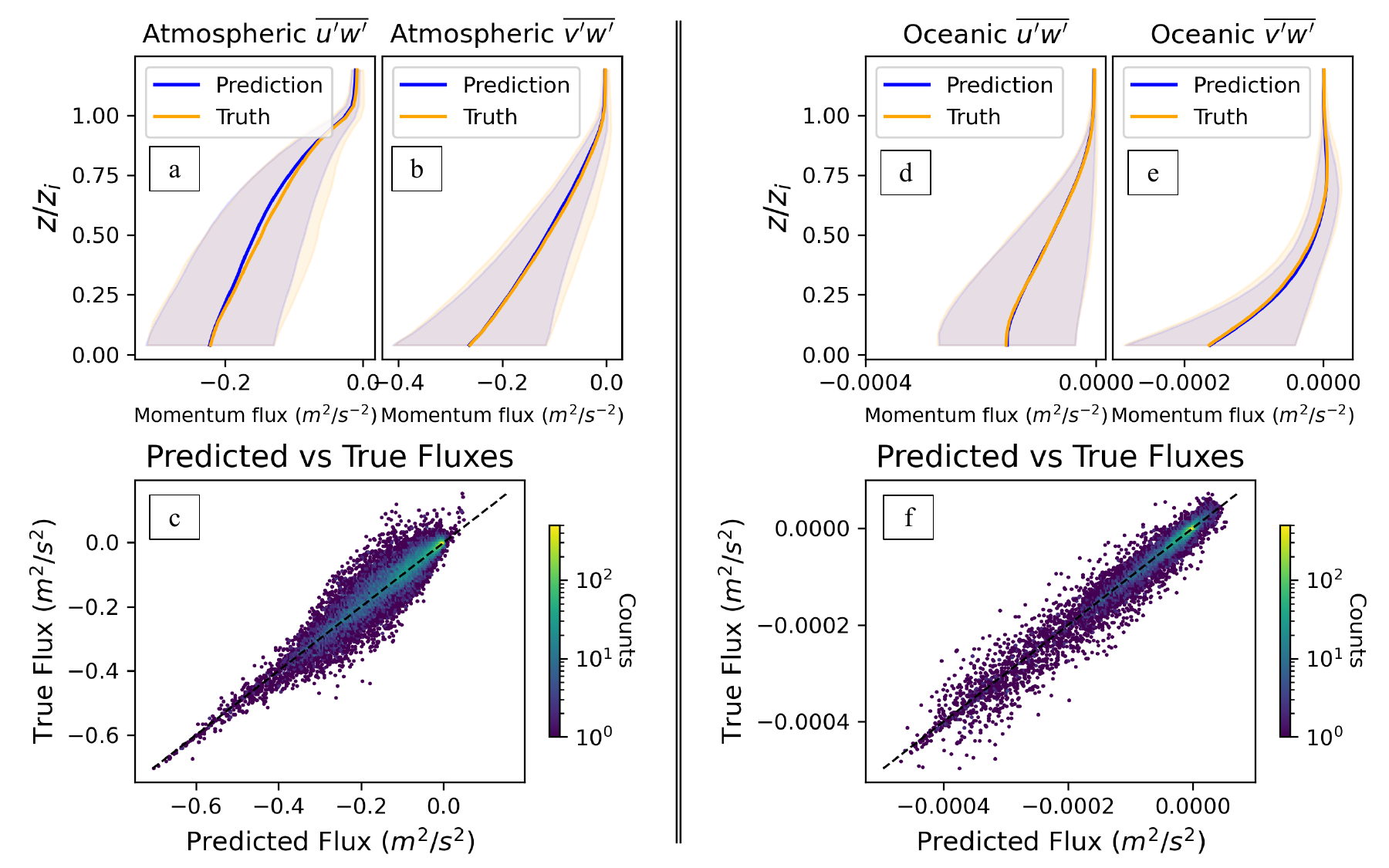}
    \caption{Offline predictions of the testing set. The testing samples were split between the atmospheric (left) and oceanic (right) features. The top panels (a,b, d, e) show the average true vertical profiles of momentum fluxes in $m^2/s^2$(orange thick lines) and average ANN predictions (thick blue lines). The shaded areas correspond to the standard deviations of the samples. The unnormalized fluxes are shown here. The bottom panels (c, f) show the density plots of predicted versus true fluxes, the values at each vertical level is considered for every test sample. }
    \label{fig:fig2}
\end{figure}

    Initially, the 9,924 samples are randomly split into training (80\%), validation (10\%) and testing (10\%) datasets. The testing set thus includes both oceanic and atmospheric samples. The ANN prediction skills are assessed by computing the prediction $R^2$, defined as:
\(
R^2 = 1 - \frac{\sum_{i=1}^n \left( y_i - \hat{y}_i \right)^2}{\sum_{i=1}^n \left( y_i - \bar{y} \right)^2}
\), where $\hat{y}_i$ are the predicted flux values, $y_i$ are the true values, and $\overline{y}$ the mean of the true fluxes. The prediction skills of both atmospheric and oceanic test samples are very good (Fig \ref{fig:fig2}), with an overall $R^2$ value of 0.94 and 0.93, respectively. Here, the $R^2$ are computed by considering the values of the fluxes at each vertical level, and for both components $\overline{u'w'}$ and $\overline{v'w'}$ jointly. The mean vertical profiles of ANN-predicted fluxes closely match the mean true fluxes in both fluids (Fig. \ref{fig:fig2} a, b, d and e, thick lines), and the ANN also captures the variability of the fluxes in the testing set (shaded areas). Note that the unnormalized fluxes are shown here (i.e., after multiplying each predicted flux by its surface value). 

Since surface values are computed by other components of GCMs and provided to the BL schemes as inputs, we first consider that there is no error on the surface value and use the true LES surface values to unnormalize the fluxes. The prediction thus always matches the true flux at the surface. However, in realistic simulations, errors might exist in the prediction of the surface momentum flux values by surface parameterizations used in models. In order to ensure that the prediction skills of the model are not significantly degraded when the prediction of the surface value is biased, the $R^2$ were recomputed after applying biases to the surface values used to unnormalize the predictions. In CAM, biases on the surface values are usually relatively small compared to observations, but can reach extreme values of about 30\% in certain cases \cite{lindvallEvaluationNearSurfaceParameters2013,burnsComparisonDielCycle2018}. Therefore, biases ranging from -30\% to +30\% (in 5\% increments) have been applied to the predictions before recomputing the $R^2$. It is found that even with largest biases (plus or minus 30\%), the overall prediction is not significantly degraded, and the prediction $R^2$ remains above 0.80 (Fig. S1).

\subsection{ANN captures upgradient fluxes}
    As mentioned in the introduction, one of the setbacks of first-order parameterizations is that by definition, they are unable to capture upgradient fluxes \cite{helferRoleShallowConvection2021}. However, in realistic situations, upgradient fluxes exist, usually arising from mesoscale dynamics, low-level jets and non-local transport \cite{larsonMomentumTransportShallow2019,dixitCounterGradientMomentumTransport2021}. Upgradient fluxes occur when the vertical gradient of mean wind or current is of the same sign as the momentum flux. In that case, turbulence acts to amplify vertical gradients rather than reducing them. Counter-gradient momentum fluxes shape the vertical profiles of wind and constitute a substantial part of the momentum transport in organized convection \cite{Savazzi2023Convection}. They also have a strong impact on jet dynamics, including their location, formation and sustainability \cite{farrellStructureSpacingJets2007,birnerUpgradientEddyFluxes2013,dixitCounterGradientMomentumTransport2021}. Capturing these fluxes is therefore essential to accurately model these dynamics.
    
    Examples of such upgradient fluxes can be found in the testing set. We define upgradient fluxes as cases where, within a BL segment, the momentum flux ($\overline{u’w’}$ or $\overline{v’w’}$) has the same sign as the vertical gradient of the corresponding mean flow component ($\overline{u}$ or $\overline{v}$). For a sample to be counted as upgradient, this criteria must occur over a continuous layer at least 100 meters deep in the atmosphere or 10 meters deep in the ocean. To avoid near-zero values, only fluxes and gradients exceeding 5\% of their respective maximum magnitudes within the upgradient layer are considered. Therefore, an atmospheric sample ($\overline{x}$, $\overline{x'w'}$), where $x$ can be either $u$ or $v$, is considered upgradient is there exists a layer of at least a hundred meters deep where:

    $\overline{x'w'}_{layer} > 0.05*Max(\overline{x'w'})$ and $(\frac{\partial{\overline{x}}} {\partial z})_{layer} > 0.05*Max(\frac{\partial{\overline{x}}} {\partial z})$
    
    or 
    
    $\overline{x'w'}_{layer} < -0.05*Max(\overline{x'w'})$ and $(\frac{\partial{\overline{x}}} {\partial z})_{layer} < -0.05*Max(\frac{\partial{\overline{x}}} {\partial z})$.

    For ocean samples, this criterion must be verified within a layer of at least 10 meters deep. It is found that among 1,984 vertical profiles in the testing set (992 samples, considering both components $\overline{u}$ and $\overline{v}$), 210 of them contained upgradient fluxes, or approximately 10\%. Out of the 210 upgradient fluxes identified in the testing set, it is found that the ANN also predicts an upgradient flux in 199 of them (95\%), with an average RMSE of 0.040 $m^2/s^2$ for atmospheric samples and 1.33.$10^{-5} m^2/s^2$ for oceanic samples.
    
\begin{figure}[h]
        \centering
        \includegraphics[width=1\linewidth]{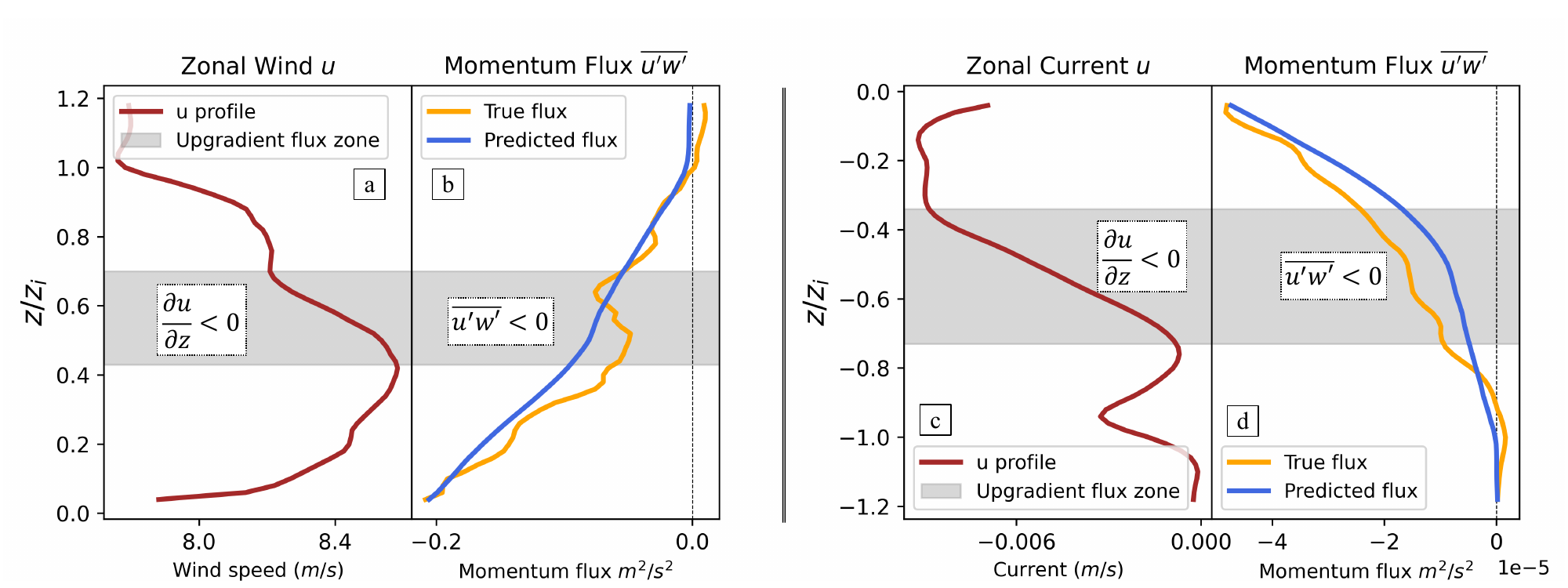}
        \caption{Upgradient momentum fluxes from the testing set. Two different samples of mean zonal wind (a) and current (c) are shown, as well as the corresponding momentum fluxes (b, d, thick orange lines) computed from the LES. The ANN-predicted momentum fluxes are also shown (b, d, thick blue lines). The gray shaded areas represent the parts of the boundary layers where the fluxes are upgradient.}
        \label{fig:fig3}
    \end{figure}

    Examples of such upgradient fluxes captured by the ANN are shown on Fig. \ref{fig:fig3}. In the atmospheric example (Fig. \ref{fig:fig3}, a and b), the vertical gradient of mean wind $u$ is negative from about $z/z_i$=0.42 to $z/z_i$=0.70 . In that layer, the momentum flux is also negative, i.e., the momentum flux is upgradient. In the oceanic example, both the vertical gradient of mean current and momentum flux are negative between about $z/z_i$=-0.35 to $z/z_i$=-0.70, resulting in upgradient fluxes. In both atmospheric and oceanic cases, the ANN manages to capture the upgradient fluxes, and the predicted and true fluxes closely match, with root mean squared errors (RMSE) of 0.017 $m^2/s^2$ and 4.97.$10^{-6}  m^2/s^2$, respectively. The results presented here therefore highlight potential significant improvements in comparison to eddy-diffusivity approaches.

\subsection{On the benefits of the unified framework}
As previously mentioned, having a single parameterization for both boundary layers can provide a unified physically consistent framework for the climate modeling community. An important subsequent consideration is whether the machine learning model's performance can also benefit from this unified framework. To test if it is the case, an ocean data-limited regime augmented with atmospheric data is replicated. Specifically, two different ANNs are trained using two different training sets.

\begin{figure}[ht]
    \centering
    \includegraphics[width=1\linewidth]{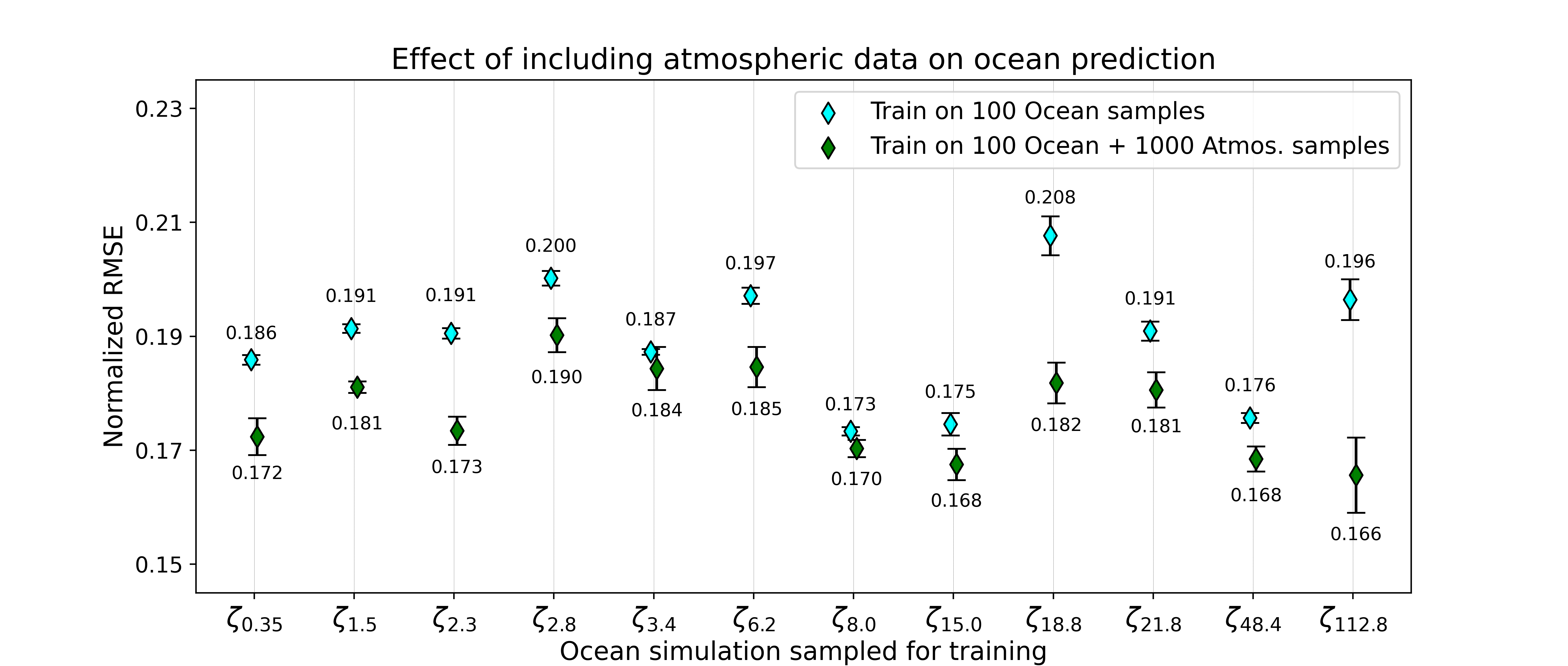}
    \caption{Improvement of the normalized RMSE after inclusion of atmospheric samples, in an ocean-data limited regime. A first ANN is trained using 100 oceanic samples randomly picked in one of the twelve LES runs, and tested on a different set of 100 oceanic samples after computing the prediction RMSE (blue diamonds). The testing samples are randomly picked in the remaining eleven LES runs. A second ANN is trained from scratch after adding 1000 atmospheric samples to the initial training set, and tested on the same initial testing set (green diamonds). The uncertainty induced by the random weight initialization of the ANN is computed by training ten separate ANNs with the same samples, and computing the mean (diamonds) and standard deviations (error bars) of the normalized RMSE. The operation is repeated for the twelve LES runs. Here, the oceanic samples change at each iteration (in both the training and testing set), but the added 1000 atmospheric samples are kept unchanged.}
    \label{fig:fig4}
\end{figure}

    Initially, 100 oceanic samples are randomly picked within one of the twelve ocean LES runs and used to train the first neural network. Another hundred oceanic samples are randomly picked from the remaining eleven LES runs and used to test the prediction accuracy of the ANN. Once this is done, a second ANN is trained from scratch after adding 1000 atmospheric samples to the initial training set (which already includes 100 oceanic samples). The prediction accuracy of this second ANN is tested on the same testing set as the first one, which contained only ocean samples. The operation is repeated for each of the twelve LES runs, such that each one of the ocean runs is sampled once for training. At each iteration, the 1000 atmospheric samples remain unchanged. For most iterations, it is found that adding atmospheric features improves the performance of the ANN (Fig. \ref{fig:fig4}). On average, the normalized RMSE is reduced by about 8\% due to this data augmentation, and reaches 16\% when training samples are taken out of the simulation $\zeta_{112.8}$.

It is essential to note that the BL wind and current profiles are not exactly similar, as the directional shear is more important in the ocean. Therefore, the inclusion of TKE as an input was essential in order for this data augmentation to significantly improve the prediction score, as TKE exhibits important similarities across fluids. This highlights the importance of carefully selecting predictive input features and performing adequate normalization procedures to build an effective fluid-independent model. These results suggest that given the similarity of the physical processes in the atmospheric and oceanic BLs, a model can benefit from any additional sample independently of the fluid it comes from, if care has been taken to ensure that the inputs and outputs are similar across fluids. In particular, one can leverage data from a fluid to enhance performance in another fluid, which is particularly interesting in a data-limited regime, where reliable model outputs and observations are scarce. In Geophysics, this often occurs because oceanic observations are typically limited, while atmospheric observations are much more abundant. The ocean parameterization might therefore benefit from the atmospheric part, in this kind of data-limited regime.

\section{Implementation in SCAM}
\subsection{Assessment of the parameterization skills}
The ANN parameterization was tested online after implementation in SCAM. For each one of the fifteen LES runs, three SCAM simulations are conducted: one with the ANN trained on all of the LES runs, one with another ANN trained from scratch after excluding the replicated LES run from the training process, and one with CLUBB, which acts as the baseline. The motivation of the second ANN run is to test the parameterization extrapolation efficiency and verify that it would still perform well on regimes excluded from the training set. SCAM is therefore run 45 times initially (fifteen LES simulations, three different parameterizations). For each one of these runs, the parameterization accuracy is assessed by comparing the SCAM-simulated wind vectors with the LES wind vectors. The normalized distance between wind vectors $D$ is computed at each vertical level and each time step, following:

$D =\sqrt{(u_{LES} - u_{SCAM})^2  + (v_{LES} - v_{SCAM})^2} /  \mid  \mathbf{U}_{LES} \mid$,\\ where $\mid \mathbf{U}_{LES} \mid= {\sqrt{u_{m,LES}^2 + v_{m,LES}^2}}$ corresponds to the average wind speed for that specific LES run.
    
\begin{figure}[h]
    \centering
    \includegraphics[width=1\linewidth]{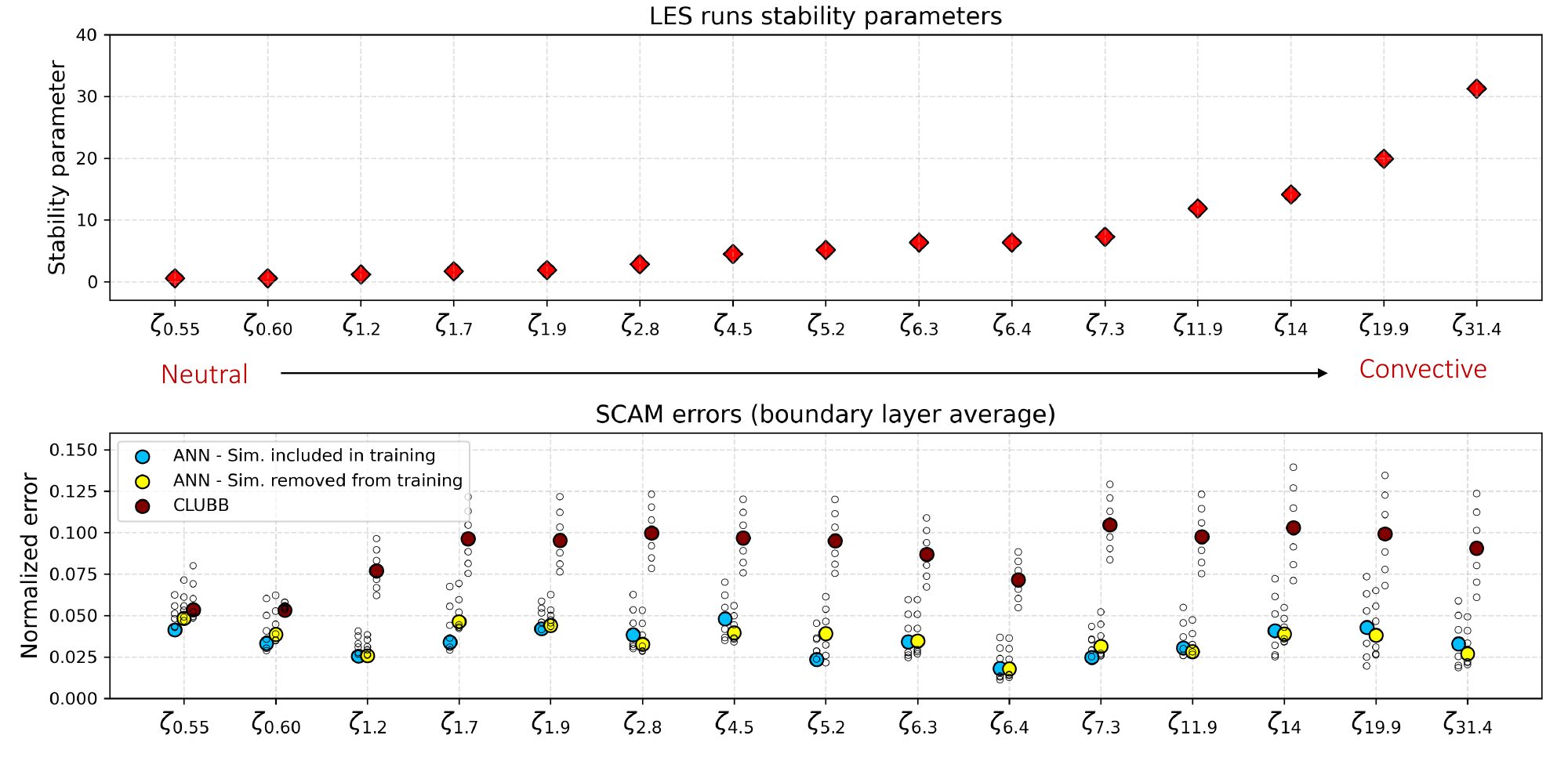}
    \caption{Stability parameters (top panel) and parameterization errors (bottom panel) for all of the fifteen replicated LES runs. The run names on the x axis correspond to the values of geostrophic wind forcing $Ug$ and surface heating $Q$ used in the simulations, and are classified in increasing order of convective strength from left to right. The error metric used is the normalized distance between the wind vectors simulated by SCAM and the LES wind vectors, averaged over every vertical levels and time steps. The blue and yellow dots correspond to the average error computed after running SCAM with the ANN parameterizations, with the replicated LES included and removed from training, respectively. The brown dots correspond to the average error of the baseline parameterization CLUBB. The small transparent dots show the errors computed after applying biases (ranging from -30\% to +30\%) to the surface flux values during the SCAM runs.}
    \label{fig:fig5}
\end{figure}

The ANN parameterization is stable for the two hours of simulation and outperforms CLUBB for each one of the fifteen LES run (Fig. \ref{fig:fig5}, bottom panel), with average normalized errors consistently lower in the ANN runs (blue and yellow dots) than in the CLUBB runs (brown dots). Here, the normalized errors are averaged over all vertical levels and over the two hour length of simulation. For all of the fifteen replicated LES runs, the errors are almost identical for the two ANN parameterizations (i.e., whether the replicated simulation is included in the training process or not). This result confirms that the relatively good skill of the ANN parameterization is not due to having seen the replicated profiles during training, which also implies that the model is not overfitting and generalizes well to different turbulent regimes. 

The different LES runs are classified according to their stability parameter (increasing convective strength going from left to right). Only the most neutral simulation ($Ug_{25}Q_{0.02}$) shows relatively similar errors for both the ANN and CLUBB parameterizations, with average normalized errors of 0.041 and 0.048 in the ANN runs, and 0.053 in the CLUBB run. The ANN errors remain relatively constant across the different turbulent regimes, while the CLUBB errors visibly increase when convection is active, and the ANN parameterization still manages to capture the evolution of the BL wind in highly convective situations. Except for the two quasi-neutral LES runs (on the far left of the figure), the normalized error is at least divided by a factor 2 with the ANN parameterization, compared to CLUBB. The LES runs are split into three different categories of turbulent regimes according to their stability parameters: quasi-neutral ($0 <\zeta < 5$), moderately convective ($5<\zeta<10$) and highly convective ($\zeta>10$). For the quasi-neutral simulations, the ANN parameterization reduces the error by a factor 2.2 on average compared to CLUBB (considering the average errors of the two ANN runs). For the moderately and highly convective simulations, the ANN parameterization reduces the error by a factor 3.1 and 3.0, respectively.

    For all of the colored dots, the LES surface momentum fluxes are used at each time step as boundary conditions. In the ANN runs, the surface value is used to unnormalize the prediction and in the CLUBB runs, the value at the first atmospheric layer is taken to be equal to the LES flux value. The smaller transparent dots on the figure show the average errors computed after applying six additional biases to the "true" LES surface values during the SCAM runs (-30\%, -20\%, -10\% , +10\%, +20\% and +30\%). The ANN parameterizations are thus compared to CLUBB a total of 210 times (fifteen LES runs replicated, two different ANN parameterizations and seven surface value configurations per parameterization). Considering all of the SCAM runs conducted, the ANN parameterization outperforms CLUBB 207 out of the 210 runs. The three runs where CLUBB's errors are lower than the ANN's correspond to the quasi-neutral simulation $Ug_{25}Q_{0.02}$, when extreme negative biases on the surface value were applied. In a majority of cases, for the ANN runs, applying biases slightly increases the errors of the single column model. However for the CLUBB runs, the errors is reduced in the convective runs when a negative bias is added to the surface momentum flux values. The reason for that error reduction comes from the fact that in convective situations, the vertical profiles of wind exhibit a strong gradient near the surface, but almost zero gradient in the mixed layer. Therefore, being a down-gradient approach, CLUBB predicts an important near-surface momentum flux, along with a quasi-zero momentum flux in the mixed layer, resulting in a strong vertical gradient of momentum flux near the surface (see \ref{in-depth}). This unrealistically strong gradient induces the surface wind to decelerate too fast, and reducing this vertical gradient by adding negative bias on the surface value yields a more realistic evolution of the wind profiles. Therefore in those cases, the error on the surface value compensates the error in the representation of the vertical gradient of momentum flux and leads to an overall error reduction.

\begin{figure}
    \centering
    \includegraphics[width=1\linewidth]{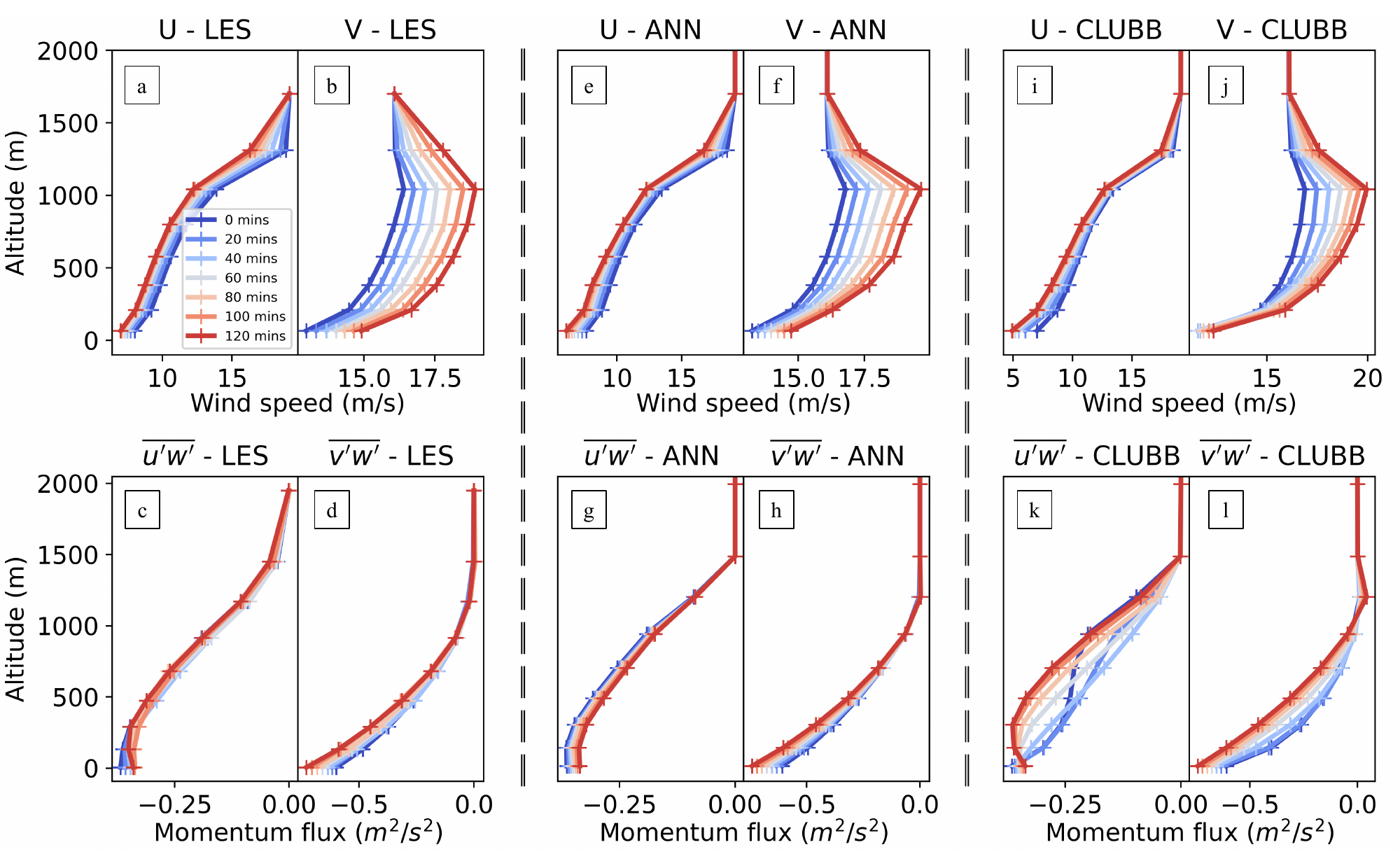}
    \caption{Temporal evolution of the boundary layer wind profiles from the LES run $Ug_{25}Q_{0.02}$ (a and b) and corresponding SCAM runs using the ANN (e and f) and CLUBB parameterizations (i and j). The corresponding vertical profiles of momentum flux computed from the LES (c and d) and predicted by the ANN (g and h) and CLUBB (k and l) are also shown. On each panel, initial vertical profiles are shown in blue and the profiles are plotted every twenty minutes.}
    \label{fig:fig6}
\end{figure}

\subsection{Boundary layer wind profiles evolution} \label{in-depth}
\begin{figure}
    \centering
    \includegraphics[width=1\linewidth]{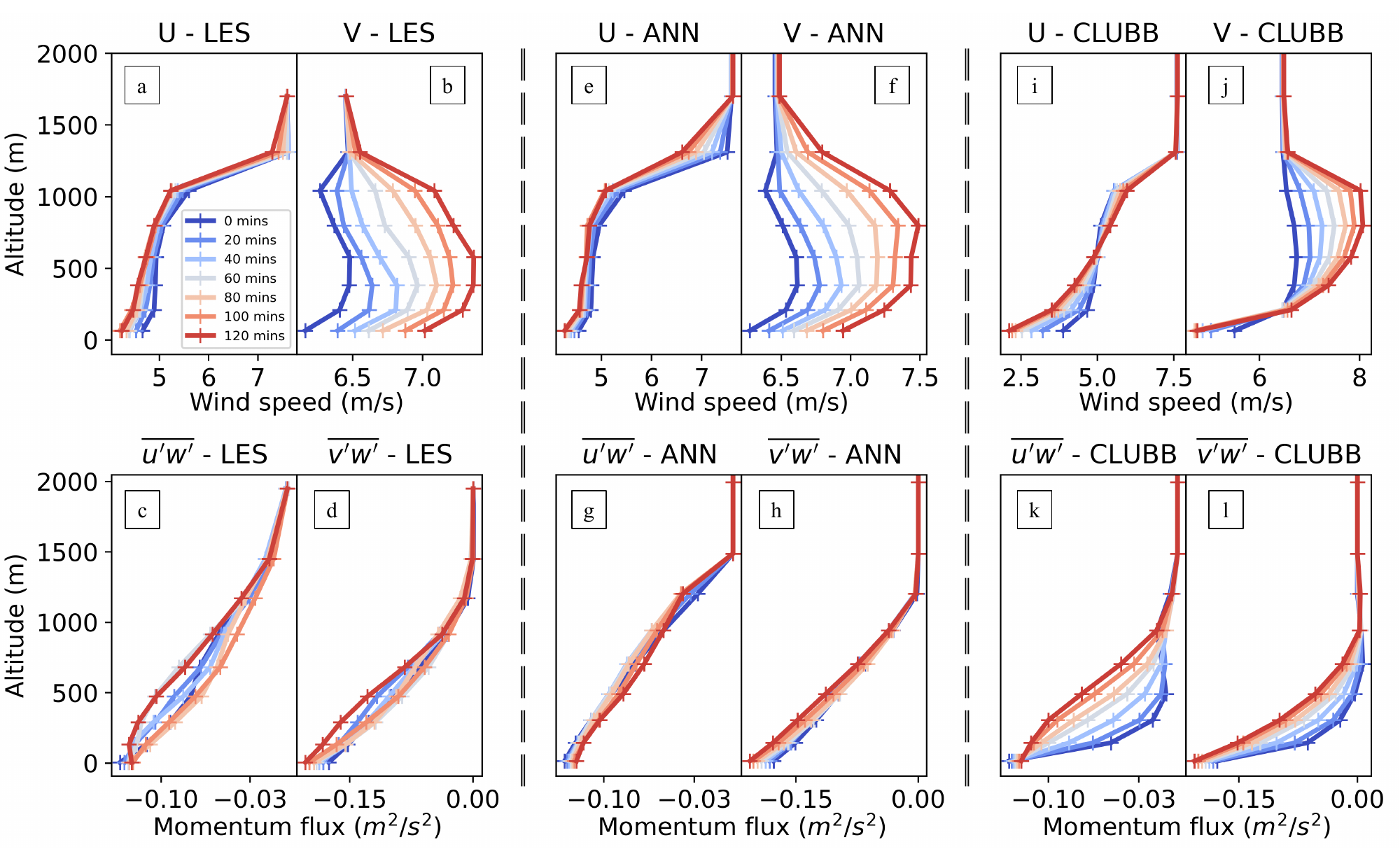}
    \caption{Temporal evolution of the boundary layer wind profiles from the LES run $Ug_{10}Q_{0.05}$ (a and b) and corresponding SCAM runs using the ANN (e and f) and CLUBB parameterizations (i and j). The corresponding vertical profiles of momentum flux computed from the LES (c and d) and predicted by the ANN (g and h) and CLUBB (k and l) are also shown. On each panel, initial vertical profiles are shown in blue and the profiles are plotted every twenty minutes.}
    \label{fig:fig7}
\end{figure}

Temporal evolution of BL wind profiles have subsequently been analyzed for two simulated cases of interest. The first case is the replicated LES simulation $Ug_{25}Q_{0.02}$, the quasi-neutral simulation that showed similar performance between the ANN parameterization and CLUBB. The second case is the replicated LES simulation $Ug_{10}Q_{0.05}$, which corresponds to the case that showed the most significant improvement when using the ANN parameterization instead of CLUBB, with the error reduced by a factor 4.2 (Fig. \ref{fig:fig5}). Doing so allows us to understand and analyze the benefits of the data-driven approach in convective situations. During the two hours of the LES runs, the zonal wind magnitude decreases while the meridional wind increases (Fig. \ref{fig:fig6} and \ref{fig:fig7}, panels a and b) due the rotation of the wind induced by the Coriolis force. In the first replicated case ($Ug_{25}Q_{0.02}$), convection is very weak and we are in a shear-driven scenario. In that case, both the ANN parameterization and CLUBB capture a correct temporal evolution of the wind profiles (Fig. \ref{fig:fig6}, e, f, i and j). In most of the boundary layer, the wind speed decreases as we get closer to the surface, and the momentum flux is downgradient (Fig. \ref{fig:fig6}, c and d), which explains why a first order approach like CLUBB performs relatively well and predict momentum flux that are close to the LES values (Fig. \ref{fig:fig6}, k and l).

In the convective run $Ug_{10}Q_{0.05}$, however, SCAM manages to predict a correct temporal evolution of the wind profiles when the ANN parameterization is active, but not when CLUBB is used (Fig. \ref{fig:fig7}, top panels). In this replicated LES, the vertical profiles exhibit characteristics of mixed layer winds, meaning that $u$ and $v$ are somewhat constant within at least part of the BL due to convective mixing. As a result of these weak vertical gradients, CLUBB predicts initial momentum fluxes that are close to zero in the mixed layer (Fig. \ref{fig:fig7}, k and l, blue lines), but stronger near the surface, which does not match the true shape of momentum flux profiles (Fig. \ref{fig:fig7}, c and d). This important near-surface vertical gradient in the momentum flux profile induces a fast decrease in the zonal and meridional wind speeds at the surface, while the speed slightly increases towards the top of the BL (Fig. \ref{fig:fig7}, i and j). After two hours of simulation, the resulting profiles (pink lines) do not exhibit the properties of a convective mixed layer, as the curves have flattened - thereby confirming the inadequacy of downgradient approaches to simulate convective BL winds. The flattening of the vertical profiles becomes evident when examining the evolution of the total horizontal wind speed profiles $U = \sqrt{u^2+v^2}$, whereas the ANN parameterization conserves a physically consistent mixed-layer wind structure (Fig. S2).

\section{Summary and Discussion}
The atmospheric and oceanic boundary layers exhibit important similarities in terms of the physical processes that shape their behavior, as well as their parameterization within climate models. Yet, no unified approach has been developed to leverage these similarities and bridge the gap between the usually separate atmosphere and ocean communities. In this study, we introduce a unified data-driven parameterization of subgrid-scale momentum fluxes for both atmospheric and oceanic convective boundary layers. The parameterization is based on an ANN trained offline on coarse-grained output from LES runs spanning a wide range of turbulent regimes in both the ocean and atmosphere. The training dataset generated includes varying combinations of surface buoyancy flux and wind forcing to represent both shear-driven and convectively-driven turbulent regimes. The ANN takes vertical profiles of mean wind for the atmosphere or current for the ocean as input and predicts the corresponding profiles of turbulent momentum fluxes. In this approach, the target fluxes are normalized by their surface values, revealing a self-similar structure across fluids and regimes that enables joint training and robust generalization to various turbulent regimes for both fluids. The unification of the two parameterizations provides a new framework with a limited number of tunable parameters, and results show that the overall model performance can be enhanced by the inclusion of features from both fluids.

The ANN demonstrates high predictive skill in offline tests, achieving R$^2$ values above 0.9 for both atmospheric and oceanic cases, and successfully capturing nontrivial features, such as upgradient fluxes—present in approximately 10\% of the testing samples—which traditional first-order schemes fail to capture. In 95\% of upgradient cases, the ANN correctly predicts the flux sign and vertical structure. Note that predicting upgradient fluxes is essential in order to simulate large-scale features such as organized convection or jet dynamics \cite{Savazzi2023Convection, farrellStructureSpacingJets2007, birnerUpgradientEddyFluxes2013, dixitCounterGradientMomentumTransport2021}, the ANN parameterization therefore shows potential in improving the representation of large-scale dynamics in climate models, which remains to be further tested online. Furthermore, the ANN maintains good performance even when the surface momentum flux used for rescaling is biased by up to ±30\%, with prediction skill (R$^2$) remaining above 0.8.

To evaluate performance in an online setting, the ANN was implemented in the Single Column Atmospheric Model (SCAM) and compared to the CLUBB parameterization. SCAM was used to replicate the wind profiles obtained from the 15 atmospheric LES runs, under different parameterization settings. To do so, the SCAM simulation was initialized with the LES profiles and forced with constant surface heating and geostrophic wind, similar to that in the LES. Results show that the ANN consistently outperforms CLUBB for all cases tested, providing substantially lower errors in reproducing the evolution of boundary layer wind profiles, especially in moderately and strongly convective regimes. For these cases, the ANN reduces the normalized wind vector error by a factor of 3 on average compared to CLUBB. Additionally, training the ANN from scratch while excluding the tested LES case confirms the model’s ability to generalize beyond its training data, i.e., to turbulent regimes that were excluded from the training set.

A detailed analysis of specific cases, which showcased either relatively low or significant improvements compared to the eddy diffusivity approach, has subsequently been conducted. Unlike first order approaches, the ANN accurately reproduces the temporal evolution of both wind and flux profiles in both shear-driven (quasi-neutral) and highly convective boundary layers, preserving physically consistent mixed-layer characteristics. These results highlight the limitations of purely downgradient approaches in convective regimes, in agreement with previous studies \cite{brownErrorsParametrizationsConvective2006, largeNonlocalTransportImplied2019a}, and demonstrate the ability of data-driven models to learn more general turbulent transport patterns from high-resolution simulations.

To the best of our knowledge, this study is the first to develop a single ANN-based momentum flux parameterization that performs effectively offline in both the atmospheric and oceanic boundary layers. Its successful implementation in SCAM suggests that such unified schemes could be implemented and potentially improve climate models. Although higher order schemes show potential to improve the representation of momentum fluxes \cite{larsonMomentumTransportShallow2019, graapUsingEUREC4ATOMIC2024}, they also add significant complexity and computation demands by adding a considerable number of prognostic equations. Mass-flux schemes provide a cheaper alternative and are able to capture some of the non-local aspect of momentum transport, but they still fail to represent some of the convective transport \cite{zhuMassFluxRepresentationVertical2015}. Data-driven approach, therefore, seems like a legitimate alternative to explore in order to improve GCM parameterizations, and the results presented here are promising steps in that direction. An interesting perspective would be to compare ANN parameterizations such as the one proposed in this study with higher order models, both in terms of skills and computation time. Although the results of this study are promising, the configuration remains idealized, and further efforts are needed to validate the parameterization in coupled GCM runs. Future work will extend this framework to include realistic and fluid-specific processes, such as condensation and precipitation in the atmosphere, or salinity, Langmuir circulation and wave-breaking in the ocean. The sensitivity of the parameterization to different vertical resolutions also has to be tested before potentially being implemented in fully coupled Earth system models under realistic boundary conditions.

\section{Appendix}

\begin{table}[hb]
    \centering
    \renewcommand{\arraystretch}{1.2}
        \begin{tabular}{|>{\centering\arraybackslash}p{2cm}|
        >{\centering\arraybackslash}p{2cm}|
        >{\centering\arraybackslash}p{2.5cm}|
        >{\centering\arraybackslash}p{2.5cm}|}
        \hline
 Name& {Geostrophic Wind  Ug (m/s)} & {Surface Heating $Q=\overline{w'\theta'}_{surf}$ (K.m/s)} & {Stability Parameter $\zeta=-z_i / L$} \\
         \hline
         Ug$_{ 25}$Q$_{0.02}$&  25&  0.02& 0.55\\
         Ug$_{16}$Q$_{0.01}$&  16&  0.01& 0.60\\
         Ug$_{12}$Q$_{0.01}$&  12&  0.01& 1.2\\
         Ug$_{16}$Q$_{0.03}$&  16&  0.03& 1.7\\
         Ug$_{20}$Q$_{0.05}$&  20&  0.05& 1.9\\
         Ug$_{16}$Q$_{0.06}$&  16&  0.06& 2.8\\
         Ug$_{16}$Q$_{0.10}$&  16&  0.10& 4.5\\
         Ug$_{12}$Q$_{0.05}$&  12&  0.05& 5.2\\
         Ug$_{8}$Q$_{0.03}$&  8&  0.03& 6.3\\
         Ug$_{5}$Q$_{0.01}$&  5&  0.01& 6.4\\
         Ug$_{10}$Q$_{0.05}$& 10& 0.05&7.3\\
         Ug$_{8}$Q$_{0.06}$& 8& 0.06&11.9\\
         Ug$_{12}$Q$_{0.20}$& 12& 0.20&14.0\\
         Ug$_{10}$Q$_{0.20}$& 10& 0.20&19.9\\
         Ug$_{8}$Q$_{0.20}$& 8& 0.20&31.4\\
         \hline
    \end{tabular}
    \caption{List of atmospheric LES runs with the geostrophic wind and surface heating forcing used. The stability parameter, defined as the opposite of the ratio of the boundary layer height $z_i$ over the Monin-Obukhov length L, is also shown.}
    \label{tab:table1}
\end{table}

\begin{table}[H]
    \centering
    \renewcommand{\arraystretch}{1.2}
        \begin{tabular}{|>{\centering\arraybackslash}p{2cm}|
        >{\centering\arraybackslash}p{2cm}|
        >{\centering\arraybackslash}p{2.5cm}|
        >{\centering\arraybackslash}p{2.5cm}|}
        \hline
 Name& {Wind Forcing 
U (m/s)}& {Surface Cooling $Q=\overline{w'\theta'}_{surf}$ (K.m/s)} & {Stability Parameter $\zeta=-z_i / L$} \\
         \hline
         U$_{15}$Q$_{-0.02}$&  15&  -0.02& 0.35\\
         U$_{15}$Q$_{-0.06}$&  15&  -0.06& 1.5\\
         U$_{8}$Q$_{-0.02}$&  8&  -0.02& 2.3\\
         U$_{10}$Q$_{-0.04}$&  10&  -0.04& 2.8\\
         U$_{15}$Q$_{-0.13}$&  15&  -0.13& 3.4\\
         U$_{10}$Q$_{-0.08}$&  10&  -0.08& 6.2\\
         U$_{8}$Q$_{-0.06}$&  8&  -0.06& 8.0\\
         U$_{10}$Q$_{-0.16}$&  10&  -0.16& 15.0\\
         U$_{8}$Q$_{-0.13}$&  8&  -0.13& 18.8\\
         U$_{5}$Q$_{-0.04}$&  5&  -0.04& 21.8\\
         U$_{5}$Q$_{-0.08}$& 5& -0.08& 48.4\\
         U$_{5}$Q$_{-0.16}$& 5& -0.16& 112.8\\
         \hline
    \end{tabular}
    \caption{List of oceanic LES runs with the surface wind and heating forcing used. The stability parameter, defined as the opposite of the ratio of the boundary layer height $z_i$ over the Monin-Obukhov length L, is also shown.}
    \label{tab:table2}
\end{table}


%
%
%
%

%



%
%

\section*{Open Research Section}
All the code used to train and test the neural network can be accessed through the following repository:
https://github.com/rflga/Boundary-layer-parameterization 

The LES data used to train the ANNs will be publicly uploaded before acceptance of the paper. For the review process, the input and output files have been uploaded as supporting information.

\section*{Conflict of Interest}
The authors declare no conflicts of interest relevant to this study.

\acknowledgments
This project is supported by Schmidt Sciences, LLC. The authors would also like to thank the M$^2$LInES team, particularly Alex Connolly, Aakash Sane, Chris Pedersen and Pavel Perezhogin for initial thoughts on the project.

\bibliography{references_BLparam}

%
%


%
%
%
%
%

\end{document}